# Conceptualising Cloud Migration Lifecycle

Mahdi Fahmideh, University of New South Wales (UNSW), Sydney, Australia
Graham Low, University of New South Wales (UNSW), Sydney, Australia
Ghassan Beydoun, University of Wollongong, Australia

*Emergent Research Forum Papers*

## 1 Introduction

Many enterprise software systems supporting IT services are characterised by a need for a high computing capability and resource consumption (Armbrust et al. 2010; Buyya et al. 2008; Koçak et al. 2013). Cloud Computing initiatives have received a significant attention as a viable solution to address these requirements through offering a wide range of services, which are universally accessible, acquirable and releasable in a dynamic fashion, and payable on the basis of service usage. Hence, organisations view the cloud services as an opportunity to empower their legacy systems.

Moving large-scale legacy systems, which may have been in operation and stored voluminous data over years, to cloud environments is not often an easy task. Previous researches acknowledge that legacy to cloud migration needs to be organised and anticipated through a methodological approach (Babar 2013; Jamshidi et al. 2013; Mohagheghi 2011; Zhao and Zhou 2014). With the guidance of a well-structured method, organisations can carry out an effective and safe application migration, instead of an ad-hoc migration which may result in a poor and erroneous migration. With respect to this, a literature review reveals that the field is abundant with many methods, techniques, decision tools, and guidelines which help to make legacy systems cloud-enabled and to exploit cloud services. Nevertheless, often these studies focus on technical aspects of legacy migration and present a different viewpoint of the same migration process. Additionally, as people in the cloud computing community may come from different backgrounds, they may use different terminologies and phrases to refer to an identical thing. Hence, it is hard to find any two migration method texts or papers which adopt the same definition of the migration process. While existing literature still have merits, the question is that how can one, easily and quickly, grasp a conceptual and technical-independent understanding of legacy to cloud migration process so that he/she can tailor it for its organisation's needs? The variety of methods which often are combined with technical considerations makes hard for to researchers and busy practitioners to digest, synthesis, and fully comprehend the migration process which is unstructured and dispersed in the existing literature. A clear need to engage more thoroughly with establishing a unified view of cloud migration process has been requested by some scholars (Hamdaqa and Tahvildari 2012; Zimmermann et al. 2012). However, to date, there is no an overarching view of such transition that reconciles multiple and disparate migration studies into an integrated and coherent model. This research stands on this idea that even though extant studies may vary in their suggested migration activities and technical details, they address the common problems of cloud migration process and are semantically identical; though, they have been expressed by different terms.

To obtain a general understanding of cloud migration process, we defined this research question: *What are the important methodological constructs (i.e. phases, activities, tasks and work-products) are typical incorporated in the process of legacy system migration to cloud environments?* Then, we have developed a generic metamodel, which integrates all domain constructs in the cloud migration literature. To represent this metamodel in a clear and well-structured manner, we used a simple version of UML (UML 2004), which is a semi-formal and de-facto for information modelling. The metamodel includes constructs in a descending order of granularity mainly *Phase*, *Activity*, *Task*, and *Work-Product*.

This research contributes to the literature in three aspects. Firstly, this is first research that uncovers and represents a generic process model of legacy system migration to cloud environments. Secondly, it can be served as a sharable repository of cloud migration knowledge which facilitates the lucid knowledge transfer about cloud migration process across and outside the cloud computing community. The metamodel, as a language infrastructure, underpins modelling, situational process model construction, and management. Finally, the proposed metamodel educates newcomers (students or novice practitioners) to the cloud computing field to envision the transition to the cloud.

The rest of this paper is structured as follows: First, Section 2 describes the background on metamodeling literature. Section 3 presents the research methodology applied to develop the target metamodel, following with presenting the resultant metamodel in Section 4. Section 5 describes a theoretical validation of the metamodel through a comparing with other existing migration model encompassing all migration activities. Finally this paper ends with a discussion on the future research plan and conclusion in Section 6.

# 2 Background and Related Work

In a simple definition a metamodel is referred to a model of a model or a model of a collection of models (Atkinson and Kuhne 2003). A metamodel for a particular domain is a specific language to describe the domain in a well-structured manner along with guidelines to specialise it into a given context at hand (Gonzalez-Perez and Henderson-Sellers 2008). It includes all related constructs, their semantics and relations in the domain. A well-designed metamodel can provide a language infrastructure to generalise, freely model the domain with abstract construct, and facilitate exchanging knowledge within the domain (Atkinson and Kuhne 2003).

Developing metamodels is a common practice in information systems researches and they have been adopted in various themes for example disaster management (Othman and Beydoun 2013) and agent-based systems (Beydoun et al. 2009), and game industry (van de Weerd et al. 2007). The common feature of these metamodels is to help users in better understanding the domain of interest. Metamodeling has been continued as an interesting topic in the cloud computing field. Some examples are the metamodel for risk identification introduced in (Keller and König 2014) and compliance and risk management for cloud services (Martens and Teuteberg 2011). Given this promising background on the adoption of metamodels, this research distils the knowledge of cloud migration process into a generic metamodel. In contrast to existing metamodel in the cloud computing field, we focused on the lifecycle of cloud migration including phases, activities, tasks, and work-products which are common in a typical cloud migration.

# 3 Research Methodology

Following the guidelines in the design science research (Hevner et al. 2004), we develop our metamodel as a specific artefact and iteratively evaluate it in the course of our research process. This paper focuses on presenting the developed metamodel as the result of the first step of this research program. As depicted in Figure 1, the development of the metamodel was conducted in two phases. The metamodel creation process started from February 2014 and finished in July 2015. In the following we summarise the metamodel creation process. A full document of these two phase is available at (Fahmideh 2015).

In the first phase, the knowledge source required for metamodeling was prepared. In order to avoid missing any useful study which could be missed by an ad-hoc literature review, we conducted a systematic literature review (SLR) to more rigorously identify approaches in the literature. On the basis of the research question was set in Section 1 and recommendations proposed by (Kitchenham et al. 2009), a SLR procedure was performed including the following steps: (i) defining search strings, (ii) selecting study sources, (iii) defining study selection criteria, (iv) conducting review, (vi) and extracting constructs, and their definitions from the identified studies. The output of the first phase was the identification of 78 studies as the source to create the metamodel.

Following the guidelines for metamodeling introduced in (Beydoun et al. 2009; Othman et al. 2014), a top-down and bottom-up iterative process was performed to create the target metamodel out of these 78 studies in the second phase. More precisely, this phase included the following steps (i) grouping similar constructs, (ii) harmonising and reconciling the constructs' definitions, (iii) organising the constructs into activities and phases in order to achieve a coherent metamodel, and finally, (iv) identifying **the relationships between the constructs.**

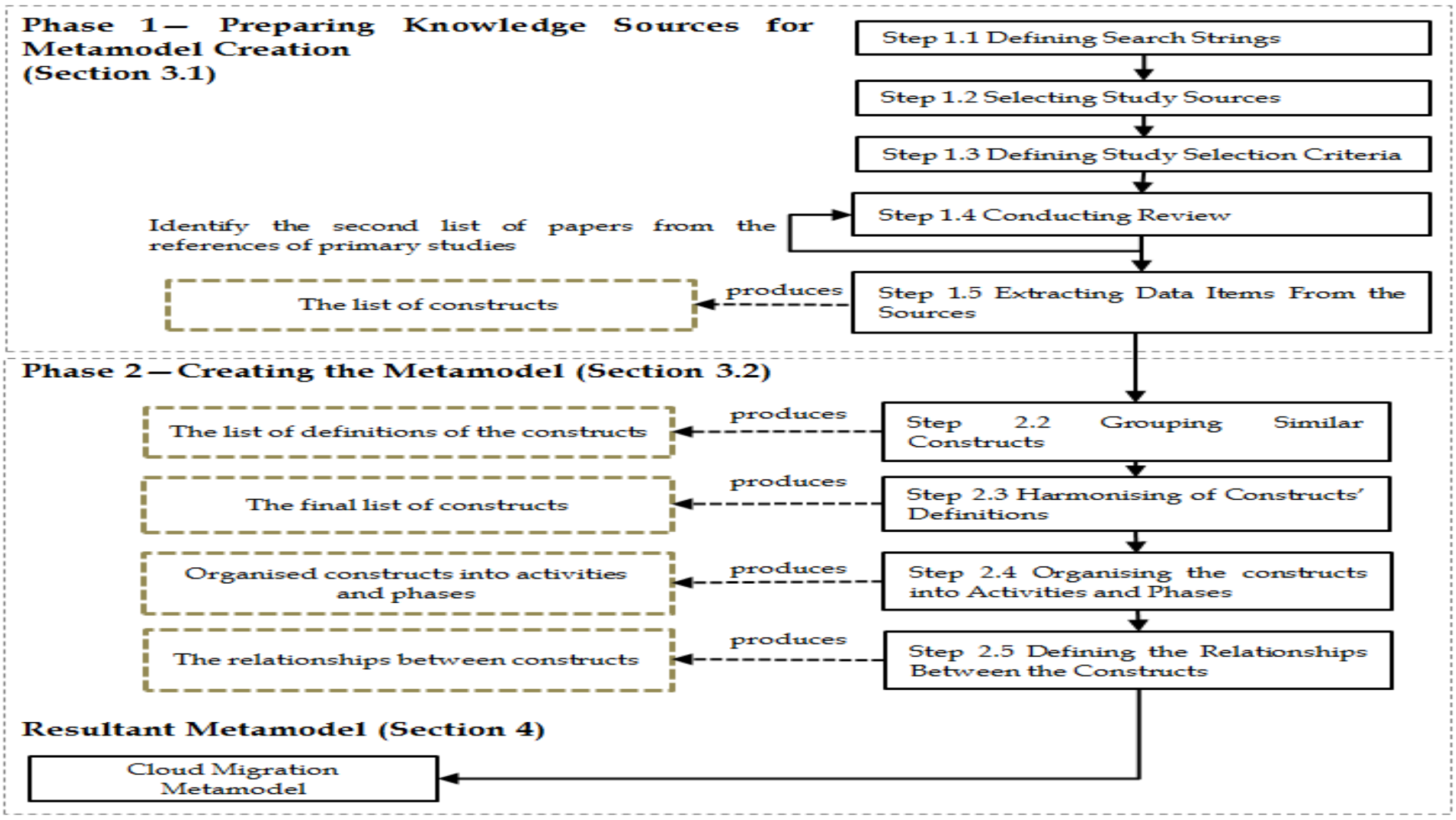

**Figure 1. Research Methodology**

# 4 The Resultant Metamodel

The derived metamodel from the literature provides a broad understanding of cloud migration process, while not narrowing down into technical and implementation details required by every situation-specific cloud migration scenario. The details are deferred to each individual instantiation of the metamodel where the users of the metamodel can adopt various implementation techniques in order to perform the tasks. This enables the gradual refinement and separation of concern during a migration process.

As shown in the diagrammatic sketches 2 to 5, the metamodel presents the classes of development constructs in the four phases of *Plan*, *Design*, *Enable*, and *Maintain*. Each phase encompasses a number of classes of activities and activities themselves constitute a number of tasks producing one or more tangible/intangible work-products. This fractal nature of the metamodel allows expressing cloud migration at multi-levels in a way that lower level adds detail to the super levels. A full description of the metamodel constructs can be found at (Fahmideh 2015).

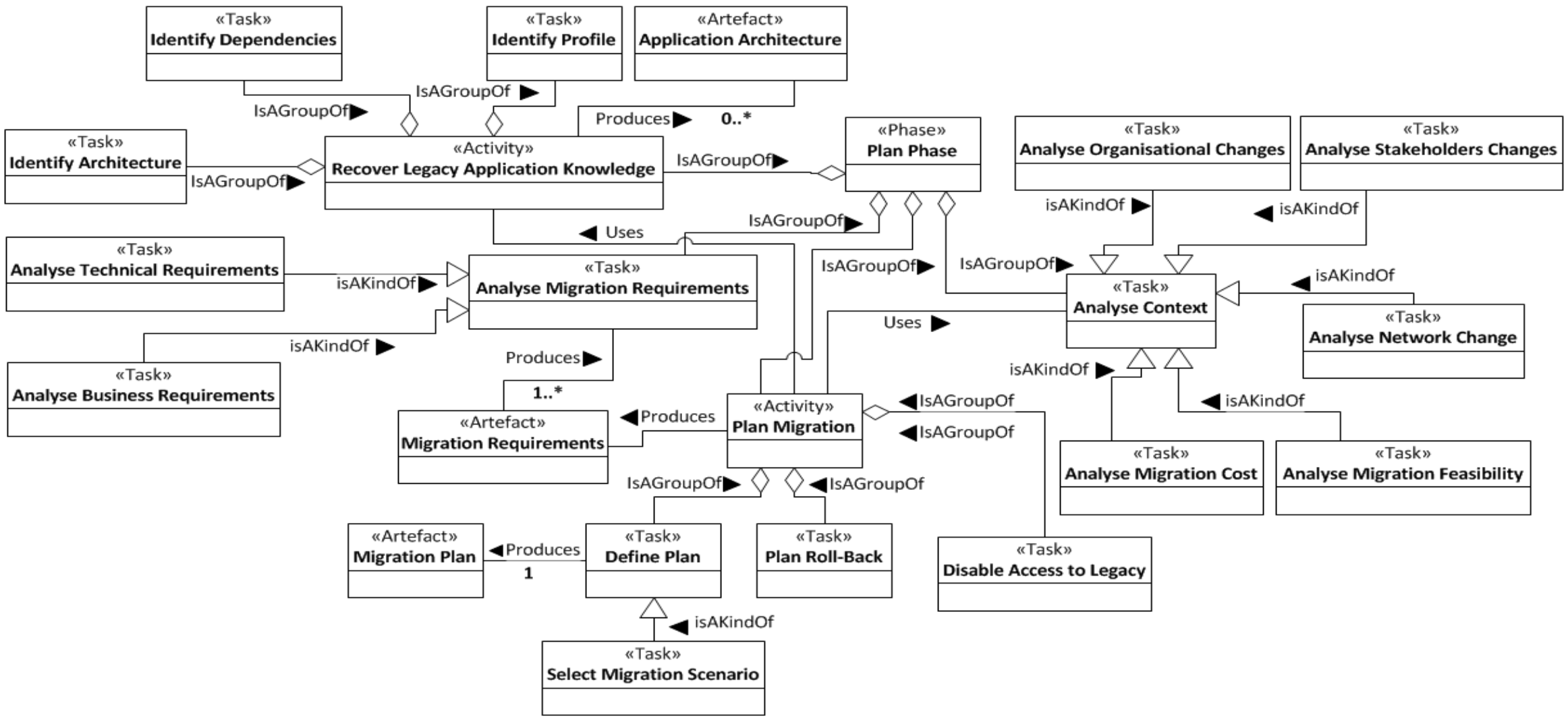

**Figure 2. Plan phase class of constructs**

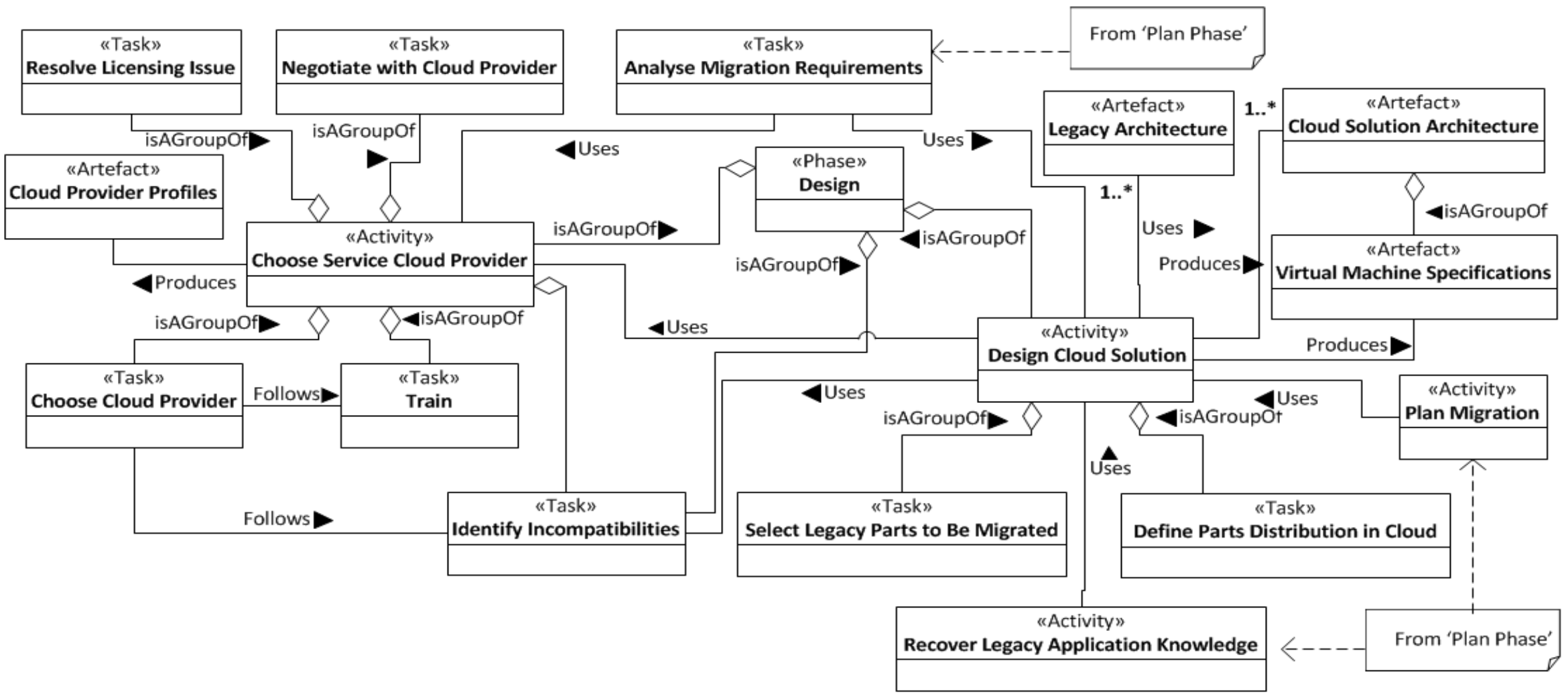

**Figure 3. Design phase class of constructs**

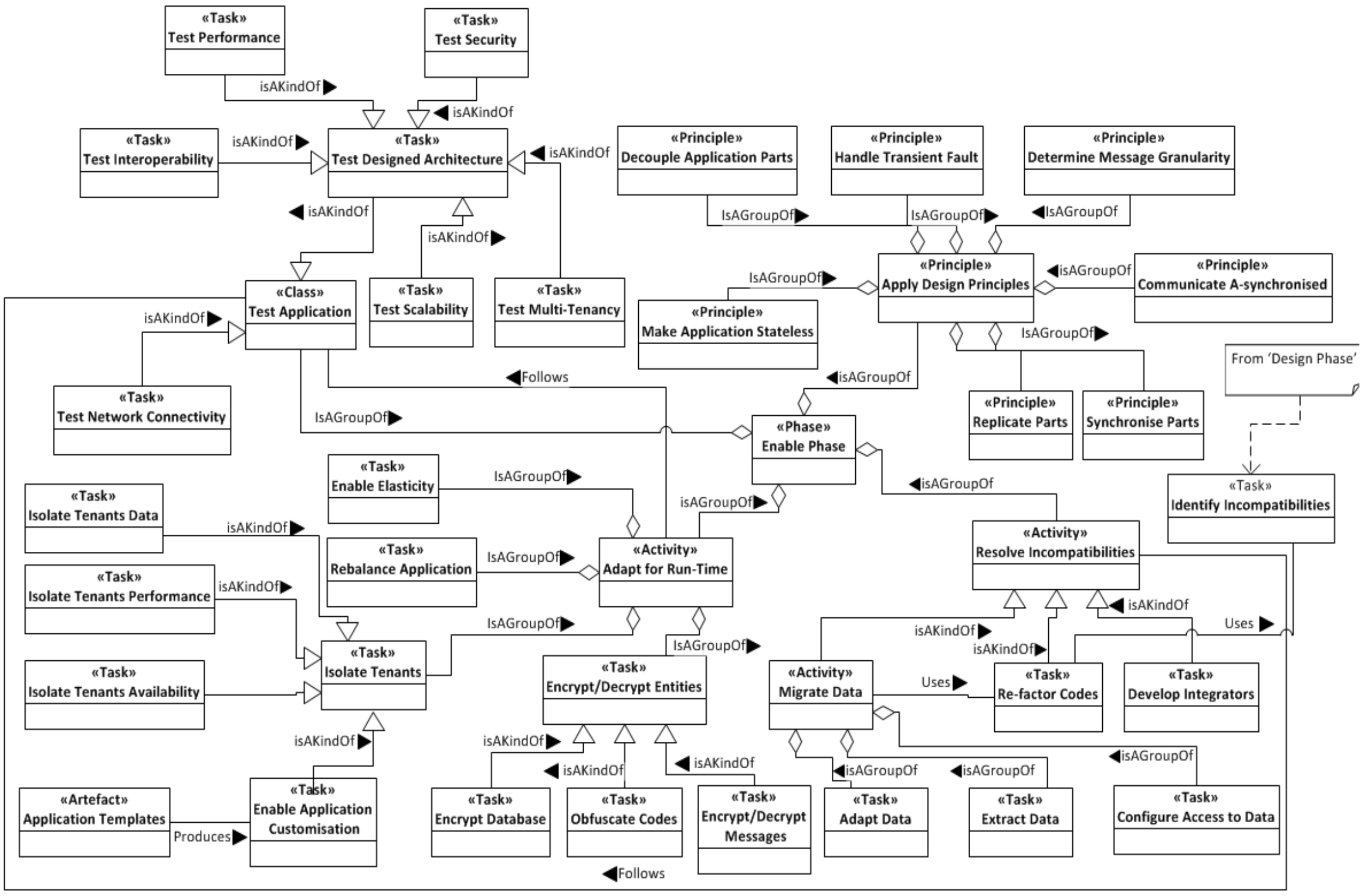

**Figure 4. Enable phase class of constructs**

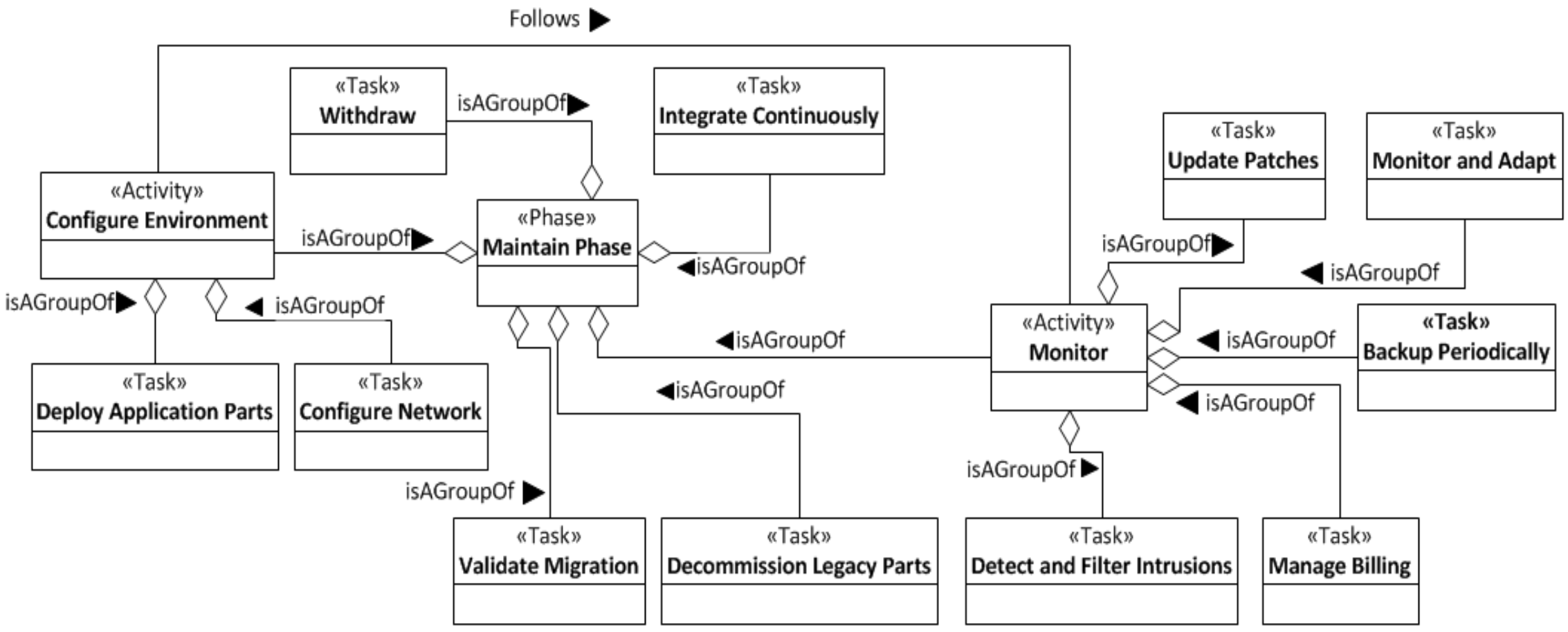

**Figure 5. Maintain phase classes of constructs**

# 5 Theoretical Validation

It is expected that a proposed metamodel should have enough generality, expressiveness and completeness to represent the domain of interest. Several techniques are used to validate a metamodel. A first common technique, suggested by (Sargent 2005), is to compare metamodel with existing metamodel/models to show how the metamodel construct can be mapped to these models. De Kok (De Kok 2010) recommends a *frequency-based selection technique* which is based on this premise that the quality of a metamodel depends on using most common constructs in the domain. Lindland et al. define *pragmatic quality assessment* which evaluates how a model can be constructed, comprehended and modified by its audience. This is determined with many properties such as quality of diagrams or text, icons and names, layout and closeness of the model to the domain (Lindland et al. 1994). They also define *Semantic quality assessment* which is concerned with the correctness and relevancy of the model to the domain whereas completeness is to check the model makes all the constructs, as much as possible, about the domain (Lindland et al. 1994). Another important technique is *Traceability* which is to validate the metamodel applicability to represent different real world scenarios where metamodel constructs are instantiated to model a given scenario (Sargent 2005). Our plan is to use the combination of these techniques to validate and improve the metamodel, accordingly. In this paper, we used the comparison technique to demonstrate how the metamodel is comprehensive enough to generate different constructs defined in other existing cloud migration models. To this aim, we selected five examples as benchmarks to appraise the completeness of the metamodel. These cloud migration models were Cloud-RMM (Jamshidi et al. 2013) Legacy-to-Cloud Migration Horseshoe (Ahmad and Babar 2014), ASDaaS (Benfenatki et al. 2013), Espadas's model (Espadas et al. 2008), and CMotion (Binz et al. 2011). For each model, we compared the constructs defined in it and the constructs in the metamodel. The definitions of constructs appeared in the process model were checked against the metamodel to find any corresponding construct in the metamodel. We found that the metamodel supports most of the constructs in these process models. For example, in Legacy-to-Cloud Migration Horseshoe, there is a construct named as *Decision on Cloud Providers*. The metamodel supports this construct through *Cloud Provider Selection* construct. Similarly, with the construct *Migration Planning*, our proposed metamodel supports this construct by *Define Plan*. Such corresponding between the proposed metamodel and Cloud-RMM is evaluable. For instance, Cloud-RMM defines the construct *Requirement Analysis* which is covered by the metamodel through the construct *Migration Requirement Analysis*. Table 1 shows how constructs in these models are corresponded with the constructs in the metamodel. The results seem to indicate that the constructs in the metamodel are indeed able to cover the constructs defined the existing migration models. That is, the selected examples can be modelled by using the proposed metamodel. This gives an initial confidence that the metamodel is valid. However, we do not claim that the metamodel is comprehensive and still there is a possibility to extend the metamodel with new constructs as discussed in the next section.

| Study | Model's construct | The metamodel's construct |
|---|---|---|
| Legacy-to-Cloud Migration Horseshoe | Feasibility Study | Context Analysis |
| | Requirement Analysis | Analyse Technical Requirements |
| | Decision on Cloud Providers | Cloud Provider Selection |
| | Migration Strategy Development | Select Migration Scenario |
| | Legacy Architecture Description | Identify Architecture |
| | Architecture Change Implementation | Resolve Incompatibilities and |
| | Code Consistency Conformance | Identify Legacy Application Architecture |
| | Migration Planning | Define Plan |
| | Cloud-Service Architecture | Design Cloud Solution |
| Cloud-RMM | Feasibility Study | Context Analysis |
| | Requirement Analysis | Migration Requirement Analysis |
| | Decision on provider | Cloud Provider Selection |
| | Sub-system to be migrated | Application Part Selection |
| | Migration Strategies Development | Migration Planning |
| | Code Modification | Code Adaptation |
| | Architecture Recovery | Application Knowledge Recovering |
| | Data Extraction | Database Schema Transformation, Query Transformation, Data Type Conversion, Emulators Implementation |
| | Architecture Adaptation | Application Cloud Enablement |
| | Test | Application Test |
| | Deployment | Application Installation |
| | Training | Initial training |
| | Effort Estimation | Migration Cost Analysis |
| | Organization Change | Impact Analysis |
| | Distribution | Application Part Selection |
| | Multi-tenancy | Access Isolation, Fault Isolation, Runtime Customizability, User Interface Adaptation |
| | Elasticity Analysis | Dynamic Resource Provisioning |
| ASDaaS | Requirements expression | Migration Requirement Analysis |
| | Service discovery | Cloud Provider Selection |
| | Service composition | Application Part Distribution |
| | Establishing the contract between the customer and the provider | Negotiation |
| | Tests and validation | Application Test |
| | IaaS selection for application deployment | Cloud Provider Selection |
| Espadas's model | Requirements | Migration Requirement Analysis, Application Knowledge Recovering |
| | Analysis | Migration Requirement Analysis, Application Knowledge Recovering, |
| | Design | Application Part Selection, Application Part Distribution, Cloud Provider Selection |
| | Implementation | Execution Phase |
| | Testing | Application Test |
| | Application Design | Cloud Solution Design |
| | Development / Migration | Application Cloud Enablement |
| | Deployment | Environment configuration |
| | Resource Planned | Dynamic Resource Provisioning |
| | Maintenance | Post Migration Review |
| | Withdraw | Withdraw |
| | Deployment Model | Cloud Solution Architecture |
| CMotion | Runtime Adaptor | Developing Adaptor |
| | Application Model | Application Knowledge Recovering |
| | Generate and Combine Alternatives | Application Part Selection, Application Part Distribution |
| | Selection and Evaluation | Cloud Provider Selection |
| | Deployment | Environment Configuration |

**Table 1. The metamodel supports constructs in existing cloud migration models**

# 6 Future Research and Conclusion

The analytical comparison of the metamodel discussed in the previous section is the first step to validate the metamodel. We plan to get insight from experts in the cloud computing in order to refine and extend the metamodel through an iterative process. To this aim, we perform a structured on-line survey to collect experts' viewpoints to assess the completeness and perceived importance of the metamodel. More specifically, they will be asked to rate the importance of the constructs in the proposed metamodel on a 1–7 Likert scale, provide comments on the metamodel, and suggest additional constructs for consideration in the metamodel. Furthermore, while the metamodel is domain-independent and generic, we anticipate that real-world cloud migration scenarios can be modelled through the metamodel constructs. This conformity is validated through conducting a series of case studies. The results the cases are used to refine the metamodel, leading to a next version of the metamodel.

This paper introduced a generic cloud migration process model which enhanced our understanding of what such cloud migration entails from the lifecycle perspective. It acts as a language infrastructure which unifies describing the process model for moving legacy systems to the cloud environments. The metamodel constitutes constructs which are common for such transition while not providing details demanded for every cloud migration scenario. Thanks to formalism was used to represent the metamodel, it can be extended with new constructs, customised for different scenarios.

## References

Ahmad, A., and Babar, M.A. 2014. "A Framework for Architecture-Driven Migration of Legacy Systems to Cloud-Enabled Software," in: Proceedings of the WICSA 2014 Companion Volume. Sydney, Australia: ACM, pp. 1-8.

Armbrust, M., Fox, A., Griffith, R., Joseph, A.D., Katz, R., Konwinski, A., Lee, G., Patterson, D., Rabkin, A., and Stoica, I. 2010. "A View of Cloud Computing," Communications of the ACM (53:4), pp. 50-58.

Atkinson, C., and Kuhne, T. 2003. "Model-Driven Development: A Metamodeling Foundation," Software, IEEE (20:5), pp. 36-41.

Babar, M.A. 2013. "Perspectives and Reflections on Cloud Computing and Internet Technologies from Nordicloud 2012," in: Proceedings of the Second Nordic Symposium on Cloud Computing & Internet Technologies. Oslo, Norway: ACM, pp. 72-79.

Benfenatki, H., Saouli, H., Benharkat, A.-N., Ghodous, P., Kazar, O., and Amghar, Y. 2013. "Cloud Automatic Software Development," ISPE CE, pp. 40-49.

Beydoun, G., Low, G., Henderson-Sellers, B., Mouratidis, H., Gomez-Sanz, J.J., Pavon, J., and Gonzalez-Perez, C. 2009. "Faml: A Generic Metamodel for Mas Development," Software Engineering, IEEE Transactions on (35:6), pp. 841-863.

Binz, T., Leymann, F., and Schumm, D. 2011. "Cmotion: A Framework for Migration of Applications into and between Clouds," Service-Oriented Computing and Applications (SOCA), 2011 IEEE International Conference on, pp. 1-4.

Buyya, R., Yeo, C.S., and Venugopal, S. 2008. "Market-Oriented Cloud Computing: Vision, Hype, and Reality for Delivering It Services as Computing Utilities," High Performance Computing and Communications, 2008. HPCC'08. 10th IEEE International Conference on: Ieee, pp. 5-13.

De Kok, D. 2010. "Feature Selection for Fluency Ranking," Proceedings of the 6th International Natural Language Generation Conference: Association for Computational Linguistics, pp. 155-163.

Espadas, J., Concha, D., and Molina, A. 2008. "Application Development over Software-as-a-Service Platforms," Software Engineering Advances, 2008. ICSEA'08. The Third International Conference on: IEEE, pp. 97-104.

Fahmideh, M. 2015. "A Development of Metamodel for Cloud Migration Process." Available at: https://www.dropbox.com/s/yut5x3cvypoz06z/A%20Development%20of%20Metamodel%20for%20Cloud%20Migration%20Process.pdf?dl=0.

Gonzalez-Perez, C., and Henderson-Sellers, B. 2008. Metamodelling for Software Engineering. Wiley Publishing.

Hamdaqa, M., and Tahvildari, L. 2012. "Cloud Computing Uncovered: A Research Landscape," Advances in Computers (86), pp. 41-85.

Hevner, A.R., March, S.T., Park, J., and Ram, S. 2004. "Design Science in Information Systems Research," MIS quarterly (28:1), pp. 75-105.

Jamshidi, P., Ahmad, A., and Pahl, C. 2013. "Cloud Migration Research: A Systematic Review," Cloud Computing, IEEE Transactions on (PP:99), pp. 1-1.

Keller, R., and König, C. 2014. "A Reference Model to Support Risk Identification in Cloud Networks,").

Kitchenham, B., Pearl Brereton, O., Budgen, D., Turner, M., Bailey, J., and Linkman, S. 2009. "Systematic Literature Reviews in Software Engineering – a Systematic Literature Review," Information and software technology (51:1), pp. 7-15.

Koçak, S.A., Miranskyy, A., Alptekin, G.I., Bener, A.B., and Cialini, E. 2013. "The Impact of Improving Software Functionality on Environmental Sustainability," on Information and Communication Technologies), p. 95.

Lindland, O.I., Sindre, G., and Solvberg, A. 1994. "Understanding Quality in Conceptual Modeling," Software, IEEE (11:2), pp. 42-49.

Martens, B., and Teuteberg, F. 2011. "Risk and Compliance Management for Cloud Computing Services: Designing a Reference Model,").

Mohagheghi, P. 2011. "Software Engineering Challenges for Migration to the Service Cloud Paradigm: Ongoing Work in the Remics Project," Services (SERVICES), 2011 IEEE World Congress on, pp. 507-514.

Othman, S.H., and Beydoun, G. 2013. "Model-Driven Disaster Management," Information & Management (50:5), pp. 218-228.

Othman, S.H., Beydoun, G., and Sugumaran, V. 2014. "Development and Validation of a Disaster Management Metamodel (Dmm)," Information Processing & Management (50:2), pp. 235-271.

Sargent, R.G. 2005. "Verification and Validation of Simulation Models," Proceedings of the 37th conference on Winter simulation: Winter Simulation Conference, pp. 130-143.

UML, O. 2004. "2.0 Superstructure Specification," OMG, Needham).

van de Weerd, I., de Weerd, S., and Brinkkemper, S. 2007. "Developing a Reference Method for Game Production by Method Comparison," in Situational Method Engineering: Fundamentals and Experiences, J. Ralyté, S. Brinkkemper and B. Henderson-Sellers (eds.). Springer US, pp. 313-327.

Zhao, J.-F., and Zhou, J.-T. 2014. "Strategies and Methods for Cloud Migration," International Journal of Automation and Computing (11:2), pp. 143-152.

Zimmermann, O., Miksovic, C., and Küster, J.M. 2012. "Reference Architecture, Metamodel, and Modeling Principles for Architectural Knowledge Management in Information Technology Services," Journal of Systems and Software (85:9), pp. 2014-2033.